\documentclass[superscriptaddress,twocolumn,showpacs,prl,floatfix]{revtex4}

\usepackage{graphics}%
\usepackage{amssymb}

\begin{document}

\title{Dynamical scaling of the quantum Hall plateau transition}

\author{F.~Hohls}
\email{hohls@nano.uni-hannover.de}
\author{U.~Zeitler}%
\author{R.~J.~Haug}
\affiliation{%
Institut f\"ur Festk\"orperphysik, Universit\"at Hannover, Appelstr.~2, 30167
Hannover, Germany }
\author{R.~Meisels}
\author{K.~Dybko}
\altaffiliation{Permanent address: Institute of Physics,
    Polish Academy of Science, Warsaw, Poland}
\author{F.~Kuchar}
\affiliation{
    Department of Physics, University of Leoben, Franz Josef Str.
    18, 8700 Leoben, Austria}

\date{\today}

\begin{abstract}

Using different experimental techniques we examine the
dynamical scaling of the quantum Hall plateau transition in
a frequency range \mbox{$f=0.1-55$ GHz}. We present a
scheme that allows for a simultaneous scaling analysis of
these experiments and all other data in literature. We
observe a universal scaling function with an exponent
$\kappa=0.5\pm0.1$, yielding a dynamical exponent
$z=0.9\pm0.2$.

\end{abstract}


\maketitle

\newlength{\plotwidth}          
\setlength{\plotwidth}{7.5cm}

\newcommand{\sxx}{{\sigma_{xx}}}  

Phase transitions between different phases of matter are
frequently met in nature, e.g. in the system ice/water,
para/ferromagnet, normal/superconductor. The usual
classification distinguishes between first and second-order
transitions. In a first-order transition the two phases
coexist at the transition temperature, in a second-order
transition they do not. Such transitions are termed
``classical'' and occur at  non-zero temperature. Different
from these types of ``classical'' phase transitions are
quantum phase transitions. Strictly spoken, they occur only
at zero temperature~\cite{sachdev99}. However, as long as
the quantum fluctuations governing the transition dominate
the thermal fluctuations, we also can observe this quantum
phenomenon at $T>0$.

Second order quantum phase transitions occur at a critical
value of a parameter which can be, e.g., the disorder in
the metal-insulator transition of two-dimensional electron
systems at zero magnetic field or the magnetic field in the
transition between Hall plateaus in such
systems~\cite{sondhi97,huckestein95}. The latter one is the
target of the investigations presented in this paper.

Generally, when the transition is approached, the
correlation length $\xi$ of the quantum fluctuations
diverges in form of a power law $\xi\propto
|\delta|^{-\gamma}$ with $\gamma$ being the critical
exponent. For a quantum Hall system $\delta$ is the
distance from a critical energy $E_c$ which can be
identified as the center of a disorder-broadened Landau
level.

For quantum Hall systems the correlation length corresponds
to the localization length $\xi(E)$, which expresses the
typical extensions of the wavefunction at energy $E$ and is
finite at all energies but $E_c$:
\begin{equation}
    \xi(E) \propto |E-E_c|^{-\gamma}
    \label{XiExp}
\end{equation}
For an infinitely large sample at \mbox{$T=0$} K the
quantum phase transition from one quantum Hall state at
$E<E_c$ to another one at $E>E_c$ happens via a single
metallic (extended) state at the critical point $E_c$; all
other states are localized. In contrast, in a finite sample
the states with a localization length larger than the
sample size $L$ are effectively delocalized, the transition
is smoothed onto a finite energy range. Additionally, at
non-zero temperature and non-zero measuring frequency
further sources of effective delocalization come into play.

The finite size dependence of the wavefunctions was
investigated in a number of numerical calculations
\cite{aoki85,chalker88,huckestein90} for non-interacting
electrons. They confirmed localization length scaling
(Eq.~\ref{XiExp}) in quantum Hall systems with a universal
scaling exponent $\gamma = 2.35 \pm 0.03$, independent of
the disorder potential~\cite{huckestein95}. Short-range
interactions are predicted not to change the critical
exponent~\cite{lee96}. However, the effect of long-range
electron-electron interaction present in the experiments
remained unclear.

Experimentally, neither the wavefunction nor the energy $E$
are directly accessible. Instead we measure the
conductivity $\sigma_{xx}(B)$ as function of the magnetic
field $B$, observing quasi metallic behavior ($\sxx\sim
e^2/h$) near some critical field $B_c$, where the state at
the Fermi energy is extended ($\xi(B) > L$), and insulating
behavior ($\sxx \ll e^2/h$) for localized states ($\xi(B)
\ll L$). More generally, the scaling theory predicts the
conductivity tensor to follow general
functions~\cite{pruisken88}
\begin{equation}
\sigma_{ij} (B) = G (L/\xi) = G_L (L^{1/\gamma}\delta B)~. \label{scaling}
\end{equation}
where we have used Eq.~\ref{XiExp} and linearized $\delta B
= B-B_c \propto E_c - E$ near the critical point. Then the
width of the quasi metallic region $\Delta B$, called
plateau transition width, follows $\Delta B\propto
L^{-1/\gamma}$. When analyzing the $\Delta B$ as a function
of the sample size $L$ such a prediction was indeed
verified experimentally~\cite{koch91}, yielding $\gamma =
2.3\pm0.2$. Quite recently, this value was also confirmed
by indirect measurements of the localization length
$\xi(B)$ in the ``insulating'' variable range hopping
regime~\cite{hohls01b,hohls02a}.

Nonzero temperature $T>0$ or frequency $f>0$ introduce
additional time scales $\tau_T \sim \hbar/k_B T$ or
$\tau_f\sim 1/f$~\cite{sondhi97}. This time has to be
compared to the correlation time $\tau_\xi\propto \xi^z$,
which is related to the correlation length $\xi$ by the
dynamical exponent $z$. In a more descriptive approach, the
additional time scale $\tau$ can be translated into an
effective system size $L_{\rm eff}\propto \tau^{1/z}$.
Plugging this into Eq.~\ref{scaling} we find scaling
functions
\begin{eqnarray}
& & \sigma_{ij}(f,T\!=\!0)=G_f (f^\kappa \delta B)
~~~{\rm and} \nonumber\\
& & \sigma_{ij}(f\!=\!0,T) = G_T (T^\kappa \delta B) \label{scale_single}
\end{eqnarray}
with a universal scaling exponent $\kappa = 1/z\gamma$. The
plateau transition widths are then given by $\Delta B
\propto T^\kappa$ and $\Delta B\propto f^\kappa$. When
identifying $\tau$ with the phase coherence time, given by
excitations with energy $k_BT$ or $hf$, $L_{\rm eff}$ is
interpreted as the phase coherence length and given by the
diffusion law $L_{\rm eff}^2\sim D\tau$. This yields a
dynamical exponent $z=2$ validated in a numerical
calculation of the frequency dependence of $\sxx$ for
non-interacting electrons~\cite{baeker99}. While $z$ is not
affected by short-range interactions~\cite{wang00}, it has
been claimed, that long-range Coulomb interaction changes
the dynamical exponent to
$z=1$~\cite{polyakov93,huckestein99}.

Temperature scaling experiments done so far yield an
ambiguous picture. While first experiments on InGaAs
quantum wells claimed to observe scaling with a universal
exponent $\kappa=0.42\pm0.04$~\cite{wei88}, a number of
other experiments reported either scaling with a sample
dependent $\kappa$~\cite{koch91b}, or even doubted the
validity of scaling at all~\cite{shahar98}. For frequency
scaling the evidence is even worse. Engel {\sl et
al.}~\cite{engel93,shahar95c} claim to observe scaling for
two different samples with $\kappa\approx 0.42$, an
experiment of Balaban {\sl et al.}~\cite{balaban98}
contradicts scaling. In addition, all these experiments
were not performed in a parameter range where $hf \gg k_BT$
is obeyed and, therefore, any single parameter scaling
analysis can not be straightforwardly applied.

To overcome these limitations we combine in this work
experiments covering a frequency range \mbox{$0.1-55$ GHz}.
We have measured the conductivity $\sxx$ in two different
experimental setups. In a frequency range from \mbox{1 MHz}
to \mbox{6 GHz}~\cite{hohls01a,hohls01b} we use coaxial
conductors, replaced by wave-guides for higher frequencies
from \mbox{26 GHz} to \mbox{55
GHz}~\cite{kuchar00,dybko01}. Our experiments are
complemented with data from
literature~\cite{engel93,shahar95c,balaban98,lewis01phd}
and present a two-variable rescaling scheme for frequency
and temperature scaling. Using all these
data~\cite{engel93,shahar95c,balaban98,hohls01a,hohls01b,kuchar00,dybko01,lewis01phd}
we will demonstrate a universal function with a universal
scaling exponent $\kappa=1/z\gamma=0.5\pm0.1$, yielding a
dynamical scaling exponent $z=0.9\pm0.2$.

\begin{figure}  
  \begin{center}
  \resizebox{\plotwidth}{!}{\includegraphics*{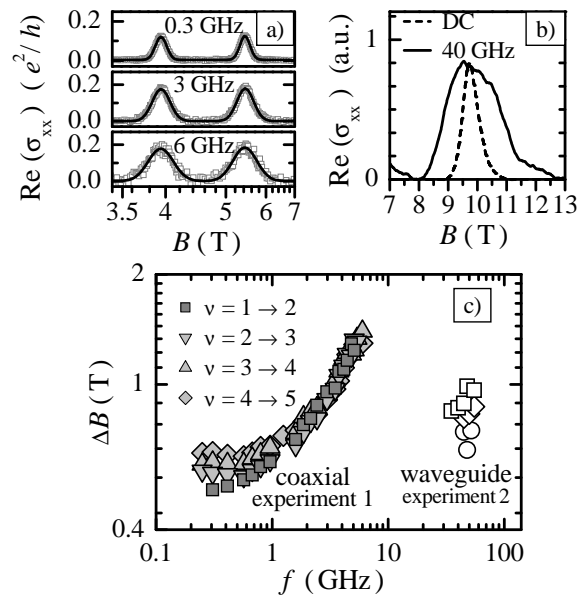}}
  \end{center}
  \caption{(a) Conductivity of sample 1 measured in the coaxial setup for different
    frequencies. Here shown are the conductivity peaks at the $\nu = 4\rightarrow
    3$ and $\nu = 3\rightarrow 2$ plateau transition.
    \protect{\newline}
    (b) Conductivity of sample
    2 at the $\nu = 2\rightarrow 1$ transition,
    measured in the waveguide reflection setup for very high frequencies
    and in a classical Hall setup for DC.
    \protect{\newline}
    (c) Frequency dependence of the width of the
    quantum Hall plateau transitions, measured for $\nu > 2$ as full width and for $\nu = 2\rightarrow1$
    as half width at half maximum~\cite{endnote1,dybko01}. For the waveguide experiment
    different symbols denote different samples from the same waver resp.\ different
    cooldown cycles with slightly different carrier densities.
    }
  \label{OurData}
\end{figure}

Frequencies up to \mbox{6 GHz} are realized in a coaxial
reflections setup as described in detail in
Refs.~\cite{hohls01a,hohls01b}. The sample, patterned into
Corbino geometry, acts as load of a coaxial cable fitted
into a $^3$He/$^4$He dilution refrigerator. The sample
conductivity is derived from reflection measurements. The
2DES used in these experiments was realized in an
AlGaAs/GaAs heterostructure with electron density
$n_e=3.3\cdot 10^{15}\;{\rm m}^{-2}$ and mobility
$\mu_e=35\;{\rm m}^2$/Vs. Traces of $\sxx(B)$ shown in
Fig.~\ref{OurData}a reveal a peak at every transition
between quantum Hall states. The spin-split transitions are
resolved up to a filling factor $\nu=n_eh/eB=6$. The
transition widths $\Delta B$ measured as the full width at
half maximum of the peaks are shown in Fig.~\ref{OurData}c
\footnote[1]{For the $\nu=1\rightarrow 2$ transition in the
lowest Landau band only the half width at the high filling
side was used to determine $\Delta B$, the critical point
$B_c$ was kept fixed at its low-frequency value. This is
necessary because an additional shoulder appears in $\sxx$
which is assigned to the presence of attractive impurities
in the sample~\cite{haug87}. Therefore, a definition of
$\Delta B$ by the full width is no more possible.}.
For \mbox{$f \leq 1$ GHz} $\Delta B$ is governed by the
temperature \mbox{$T\approx 0.1$ K} of the 2DES as deduced
from temperature dependent measurements at \mbox{$f=0.2$}
GHz. Above \mbox{2 GHz} frequency scaling $\Delta B \propto
f^\kappa$ takes over.

In a second experiment we use waveguides to access
frequencies from 26 to 55 GHz. The sample with the 2DES,
realized in a standard AlGaAs/GaAs heterojunction, is
placed at the end of the waveguide and partially reflects
the incident microwave. We modulate the carrier density
using a thin front gate to discriminate the 2DES
contributions from other reflections, for details
see~\cite{kuchar00,dybko01}. The electron density and
mobility are $n_e=3.6\cdot10^{15}\;{\rm m}^{-2}$ and
$\mu_e=10\;{\rm m}^2$/Vs. The use of a low mobility sample
ensures that $hf$ (\mbox{0.2 meV} at \mbox{50 GHz}) is
smaller than the Landau Level width \mbox{($\sim 1$ meV)}.
For higher Landau levels we do not observe a distinct spin
splitting and thus concentrate for this experiment on the
$\nu=1\rightarrow 2$ transition. The measured conductivity
$\sxx(B)$ at a temperature of \mbox{$T=0.3$ K} is shown in
Fig.~\ref{OurData}b, the evaluated transition
widths~\cite{endnote1,dybko01} are depicted in
Fig.~\ref{OurData}c.

\begin{table}[b]  
 \caption{Key data of the compared experiments: 2DES mobility $\mu_e$ in m$^2$/Vs,
 analysed range of filling factor $\nu$, 2DES \mbox{temperature $T$}, and
 \mbox{frequencies $f$}. The 2DES used in the
 experiment of Shahar {\sl et al.} resides in InGaAs, for the other Experiments AlGaAs/GaAs heterostructures
 were used.} \label{table}
  \begin{tabular}[t]{cccccc} \hline \\[-10pt]
 experiment & $\mu_e \left(\frac{\rm m^2}{\rm Vs}\right)$ & $\nu$ &
                \hspace*{3ex}$T$~(K)\hspace*{3ex}& $f\,$({\small GHz}) \\[5pt] \hline
 coaxial, exp.1~\cite{hohls01a,hohls01b}  & 35 & 1-5 & 0.1 & 0.1-6\\
 waveg., exp.2~\cite{kuchar00,dybko01}  & 10 & 1-2 & 0.3 & 35-55 \\ \hline
 Engel {\sl et al.}~\cite{engel93}  & 4 & 1-2 & 0.14-0.5 & 0.2-14\\
 Shahar {\sl et al.}~\cite{shahar95c}  & 3 & 0-1 & 0.2-0.43 & 0.2-14\\
 Balaban {\sl et al.}~\cite{balaban98}  & 3 & 1-2 & 0.15? & 0.7-7\\
 Lewis~\cite{lewis01phd}  & 50 & 3-5 & 0.24-0.5 & 1-10
 \end{tabular}
\end{table}

\begin{figure} 
  \begin{center}
  \resizebox{0.84\plotwidth}{!}{\includegraphics*{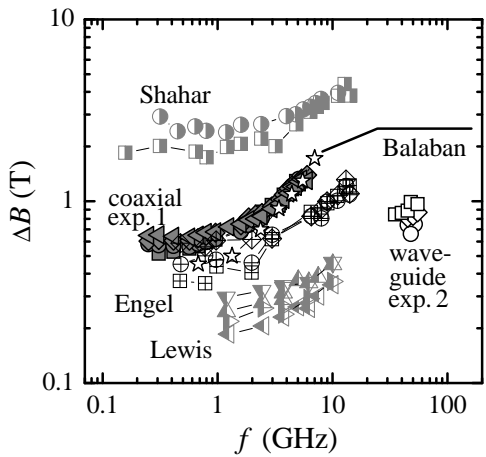}}
  \end{center}
  \caption{Comparison of the transition width vs.~ frequency for all
  experiments~\cite{engel93,shahar95c,balaban98,hohls01a,hohls01b,kuchar00,dybko01,lewis01phd}.
  Multiple symbols for the
  same experiment denote different samples or cooldowns (waveguide), transitions (coaxial,
  Lewis) or temperatures (Engel, Lewis, Shahar) as presented in
  Table~\ref{table}.
    }
  \label{AllData}
\end{figure}

The combination of our two experimental techniques allows
to investigate the scaling behavior of the quantum Hall
plateau transition in a large frequency range.
Additionally, we can compare our results to all other
frequency scaling experiments. Table~\ref{table} summarizes
the wide ranges of the parameters like frequency, mobility,
density, temperature, filling factor, and material, which
were covered by the different
experiments~\cite{engel93,shahar95c,balaban98,hohls01a,hohls01b,kuchar00,dybko01,lewis01phd}.
The observed frequency dependencies of the Hall plateau
transition width $\Delta B$ are summarized in
Fig.~\ref{AllData}.

Since most of these data do not fulfill $hf \gg k_BT$, we have to take into
account the influence of both frequency $f$ and temperature $T$ for our
analysis. Therefore, the single-parameter scaling functions for the plateau
transition (Eq.~\ref{scale_single}) are modified using a two-variable scaling
analysis~\cite[p.327]{sondhi97}
\begin{equation}
\sigma_{ij} (T,f) = G_{T,f} (T^\kappa\,\delta B,f^\kappa\,\delta B)~.
\end{equation}

\begin{figure} 
  \begin{center}
  \resizebox{0.85\plotwidth}{!}{\includegraphics*{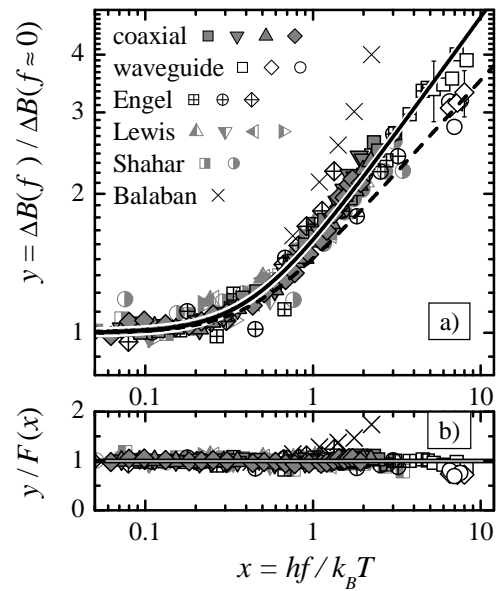}}
  \end{center}
  \caption{(a) Normalized plateau transition width
  $y=\Delta B(f,T)/\Delta B(f\approx 0,T)$ vs. dimensionless parameter
  $x=hf/k_B T$ for all data presented in Table~\ref{table} and
  Fig.~\ref{AllData}.
    The solid line results from a fit
    of $F(x)=(1+(\alpha x)^2)^{\kappa/2}$
    (Eq.~\ref{rescaleeq}) to all data except those for highest frequencies \mbox{$f>20$
    GHz}
    and the Balaban experiment ($\alpha\approx 2$, $\kappa = 0.5\pm 0.1$).
    The dashed line depicts a reduced $\kappa=0.4$.
    \protect{\newline}
    (b) Rescaling of $y = \Delta\nu(f,T) / \Delta\nu(f\approx0,T)$ by the scaling
    function $F(x)$ allows to judge the quality of scaling.
    }
  \label{Universal}
\end{figure}

Both $hf$ and $k_BT$ set an energy scale. Since frequency
and temperature act as independent processes it is a
reasonable ansatz to sum the energies
squared~\footnote{Simply adding the energies ($\Gamma =
k_BT + \alpha hf$) does not fit our data.}, resulting into
a combined energy scale \mbox{$\Gamma=[(\alpha hf)^2+(k_B
T)^2]^{1/2}$}. The factor $\alpha$ is of the order of unity
and covers the differences of the effects of frequency and
temperature. Using this simple model the transition width
scales as $\Delta B\propto \Gamma^\kappa$ and can be
rewritten as
 \begin{equation}
    \Delta B_s (T,f) =
    \Delta B_s(T)\left(\sqrt{1+\left(\frac{\alpha\,hf}{k_B T}\right)^2}\right)
     ^\kappa
    \label{rescaleeq}
 \end{equation}
with a prefactor $\Delta B_s(T)$ only depending on temperature and
on the individual sample $s$.

We use this equation to combine all data presented in
Fig.~\ref{AllData} in a single graph as shown in
Fig.~\ref{Universal}a. All the transition widths $\Delta
B(T,f)$ measured in the different experiments are
normalized to the DC width $\Delta B(T,f\approx 0)$ and
plotted versus the ratio $x=hf/k_BT$ of frequency and
temperature~\cite{engel96}. The lowest temperature $T$ of
the 2DES is estimated in the following way: Most authors
state the exponent for the temperature dependence of
$\Delta B(T)$. Combined with the low frequency width
$\Delta B$ at high temperatures where the 2DES still
couples thermally to the liquid $^3$He/$^4$He bath we
extract the lowest 2DES-temperature from the $f\rightarrow
0$ saturation width.

All data except those from Balaban {\sl et
al.}~\cite{balaban98} fall on top of our data and each
other, independent of material, mobility, density,
experimental technique, temperature and filling factor,
including the quantum Hall to insulator transition analyzed
in Ref.~\cite{shahar95c}. The single deviation observed in
Ref.~\cite{balaban98}, accompanied by deviations from
temperature scaling, is probably caused by macroscopic
inhomogeneities, which was shown to spoil any universal
scaling behavior~\cite{ruzin96}. We therefore exclude this
data from the following scaling analysis.

The observation of a universal function $F(x)$ in
Fig.~\ref{Universal} clearly reveals the universality of
the quantum phase transition between different Hall
plateaus and into the Hall insulator. We can now determine
the associated universal scaling exponent $\kappa$ by a fit
of Eq.~\ref{rescaleeq}, omitting only the highest
frequencies $f>15$~GHz from the waveguide experiment, which
will be treated separately. The fit, shown as solid line in
Fig.~\ref{Universal}, yields $\alpha\approx 2$ for the
crossover parameter between frequency and temperature and a
scaling exponent $\kappa=0.5 \pm 0.1$.

Within their individual $2\sigma$ error the data from the
second experiment using the waveguide technique also agree
with the fit function $F(x)$, but Fig.~\ref{Universal} also
reveals that their combined statistical weight indicates a
slightly lower scaling exponent $\kappa$ between $0.5$ and
$0.4$. This reduction of $\kappa$ towards its lower limit
$\kappa= 0.2$ for a non-interacting
2DES~\cite{huckestein95,wang00} possibly hints to a partial
screening of Coulomb interaction of the electrons, caused
by the gate on top of the sample.

Now that we have determined the universal scaling exponent
$\kappa=1/z\gamma=0.5\pm0.1$ of the frequency and
temperature scaling, we would like to separate the
dynamical exponent $z$ and the critical exponent $\gamma$.
Recent experiments exploited the frequency~\cite{hohls01b}
and the temperature dependence~\cite{hohls02a} of the
conductivity in the variable range hopping regime, which
allowed to determine the localization length $\xi(B)$. Both
observed a scaling behavior
$\xi\propto|\delta\nu|^{-\gamma}$ with a universal critical
exponent $\gamma = 2.3\pm0.2$ in agreement with earlier
size scaling experiments~\cite{koch91} and astoundingly
with the value obtained in numerical studies for
non-interacting
electrons~\cite{aoki85,chalker88,huckestein90}. With this
universal scaling exponent $\gamma \approx 2.3$ we can then
deduce a dynamical exponent $z=1/\gamma\kappa=0.9\pm0.2$ in
agreement with theoretical
predictions~\cite{polyakov93,huckestein99}.

In conclusion, we have developed a method to combine data
from all experiments on frequency scaling of the quantum
Hall plateau transition. We find a universal scaling
function independent of material, density, mobility,
experimental technique, temperature and filling factor,
which clearly demonstrates the universal nature of this
quantum phase transition. We determine the universal
scaling exponent of an interacting quantum Hall system to
$\kappa=0.5\pm0.1$ and, using the critical exponent $\gamma
\approx 2.3$, the dynamical exponent to $z=0.9\pm0.2$.

We thank F. Evers, B. Huckestein, B. Kramer, D.G. Polyakov,
and L. Schweitzer for useful discussions and R.M.~Lewis for
providing us with data from his PhD-thesis. This work was
supported by the DFG, DIP, and the European Union, grant
no.~FMRX-CT98-180.

\end{document}